\documentclass{aa}

\usepackage{epsfig,latexsym,longtable,lscape,supertabular}
\usepackage[authoryear]{natbib}

\def\e20{$\times 10^{20}$}
\def\ergsec{erg s$^{-1}$}

\def\today{\number\day -\number\month -\number\year}
\def\chandra{{\it Chandra}}

\begin{document}

\pagenumbering{arabic}
\title{Evidence of unrelaxed IGM around \object{IC1262} }

\author{
G. Trinchieri,  
\inst{1}
D. Breitschwerdt,
\inst{2}
W. Pietsch,
\inst{3}
J. Sulentic, 
\inst{4}
\and A. Wolter
\inst{1}}
\institute{
INAF-Osservatorio Astronomico di Brera, via Brera 28, 20121
 Milano Italy\\
 \email{ginevra.trinchieri@brera.inaf.it}
\and 
Institut f\"ur Astronomie, 
Universit\"at Wien        
T\"urkenschanzstrasse 7,
A-1180 Wien                       
Austria
\and
Max-Planck-Institut f\"ur extraterrestrische Physik,
              Giessenbachstrasse, D-85740 Garching
              Germany
\and
Department of Physics and Astronomy, University of Alabama,
 Tuscaloosa, AL 35487, USA
}

   \date{Draft: \today}
\abstract
  {}  
{A peculiar morphology of the hot gas was discovered at the center of
{IC1262} with the ROSAT HRI.  Sensitive Chandra and XMM-Newton
data were requested to
investigate the characteristics of this structure to understand its
nature.}
{We have exploited the high resolution and sensitivity of Chandra's
ACIS-S to investigate the peculiar
morphology and spectral characteristics of hot gas in the group around
{IC1262}. XMM-Newton data are only partially usable due to very heavy high
background contamination, but they are useful
to  confirm and strengthen the results from 
Chandra. }
{The Chandra data show a quite dramatic view of the {IC1262} system:
a sharp discontinuity east of the central galaxy,
with steep drops and a relatively narrow feature over 100 kpc long, 
plus an arc/loop to the N,  are all indicative of a turmoil in the
high energy component.  Their morphologies could suggest them to be 
tracers of shocked material caused
either by peculiar motions in the system or by a recent merger
process, but the spectral characteristics indicate that the structure is
cooler than its surroundings. 
The lack of evidence of significant structures in the velocity
distribution of the group members and the estimated scale of the phenomenon 
make the interpretation of its physical nature challenging. We review a
few possible interpretations, in light of similar phenomena observed in clusters
and groups.
The ram pressure stripping of a bright spiral galaxy, now near the
center of the group, is a promising  interpretation for most of
the features observed.  The relation with the radio activity requires a better
sampling of the radio parameters that can only be achieved with deeper and higher
resolution observations. }
   {}


\authorrunning{Trinchieri et al.} 
\maketitle

\section{Introduction}

The presence of hot gas in groups, which is now being detected to
fainter and fainter levels, raises many questions about the
origins of the IntraGalacticMedium (IGM) 
and its role in the formation and evolution of the whole system.
The existence of gas in groups is a strong indicator
of a common  gravitational potential, and  its detection in the X-ray
band ensures that this potential is sufficiently deep to heat gas to
X-ray emitting temperatures.  Given the short dynamical timescales,
galaxies in groups are likely to be undergoing significant interactions
and mergers.  Infall of new material, interaction and/or merging of either
accreted material or of group members can produce perturbations in the
whole system.  It is to be expected that the much smaller masses involved
than in clusters of galaxies allow the detection of small perturbations
that result from these processes in the intragroup material.  In high
resolution N-body simulations, arc-like structures with higher X--ray
surface brightness relative to the underlying smooth gas emission are
expected, since the X--ray emissivity is higher in high density regions.
Moreover, in groups even fainter structures 
will be visible than in rich clusters, because the background IGM is 
cooler and less luminous, and therefore the X-ray contrast will be better.

IC1262 was identified with a ROSAT All Sky Survey source, and selected for
follow up observations because of its unusually high X-ray to optical
flux ratio. 
The detailed ROSAT-HRI observations and optical
data have in fact suggested that the emission can be attributed to the
presence of a small group around the galaxy, which explains to first
order both the high luminosity ($L\rm _x$ $\sim$ 2.2$\times 10^{43} $ \ergsec, and
$\rm L_x / L_B \sim$ 32.7 [log (erg s$^{-1}$ L$_{\sun}^{-1}$)]) and the extension of the source
\citep[$>$ 250 kpc,\footnote{We have rescaled the original values to our
current assumption of  the   
 distance,   140 Mpc (H$\rm _o$ = 70 km s$^{-1}$), which
corresponds to a scale of $\sim 1.5''$ kpc$^{-1}$.}][]{trinchieriandpietsch2000}.  However the high resolution
image showed more than
just what might be expected from a small group.
A central complex structure, with significant surface brightness
discontinuities, was evident in the X-ray image.  This structure
appears as a perturbation over a relatively flat plateau, outside of
which there is a relatively smooth surface brightness decline, similar
to what is observed in many other groups (and usually modeled with a
$\beta$-type function).  Its luminosity, above the local emission,
account for about 10\% of the total, and it is 
comparable itself to that of small groups.

\begin{figure*}
\psfig{figure=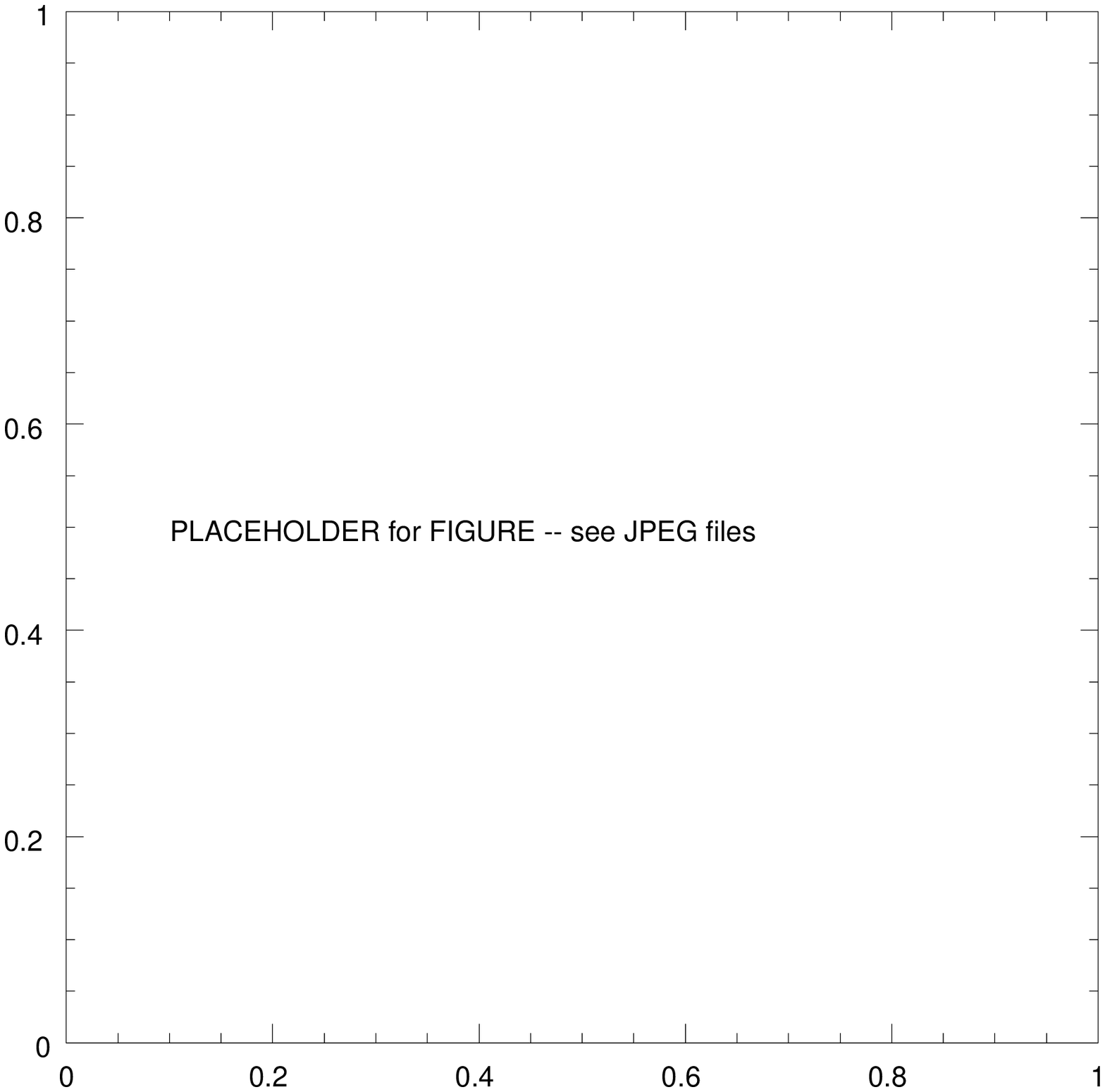,width=18cm}
\caption{Raw data (LEFT) and adaptively smoothed X-ray contours on the
blue print of the DSS2 (RIGHT) from the Chandra observation. 
Both X-ray images are in the 0.3-7.0 keV
band. The box indicates the
edges of the ACIS-S CCD7 field. CCD6 is to the SW  and CCD8 to the NE.}
\label{rawdata} 
\end{figure*}

We have obtained follow-up observations with both Chandra and XMM-Newton,
to better characterize the properties of the group and of the central
morphology and understand its nature.   Preliminary results from the Chandra data 
were presented in \citet{gt2004} and \citet{trinchierisaporiti2004}; here we 
present the detailed analysis of both sets of data. 

\citet{Hudsonetal} and \citet{Hudson} have also studied this source with BeppoSAX and have 
used the same Chandra data as a follow up of the comparison between  X-ray and
the radio emissions, prompted by the results of the BeppoSAX observation.  These
authors claim a detection of diffuse, non-thermal emission to the S of IC1262, 
which they associate with a
mini-halo in the radio, most likely produced by a merging episode with a smaller
sub-clump.   As will become clear in the data analysis and discussion sections, 
the Chandra data confirm the very complex morphology 
highlighted by the ROSAT HRI data, but also show 
a very complex temperature structure in the central region, that brings us to
somewhat different conclusions regarding the presence of non-thermal emission.
However, the paucity of data at other wavelengths is the major obstacle in our
understanding of the characteristics of this source.   We are planning to
investigate its multiwavelength properties with new data.  In the present work we
will make use of the data already available in the literature and in the public
archives. 

\begin{figure*}
\psfig{figure=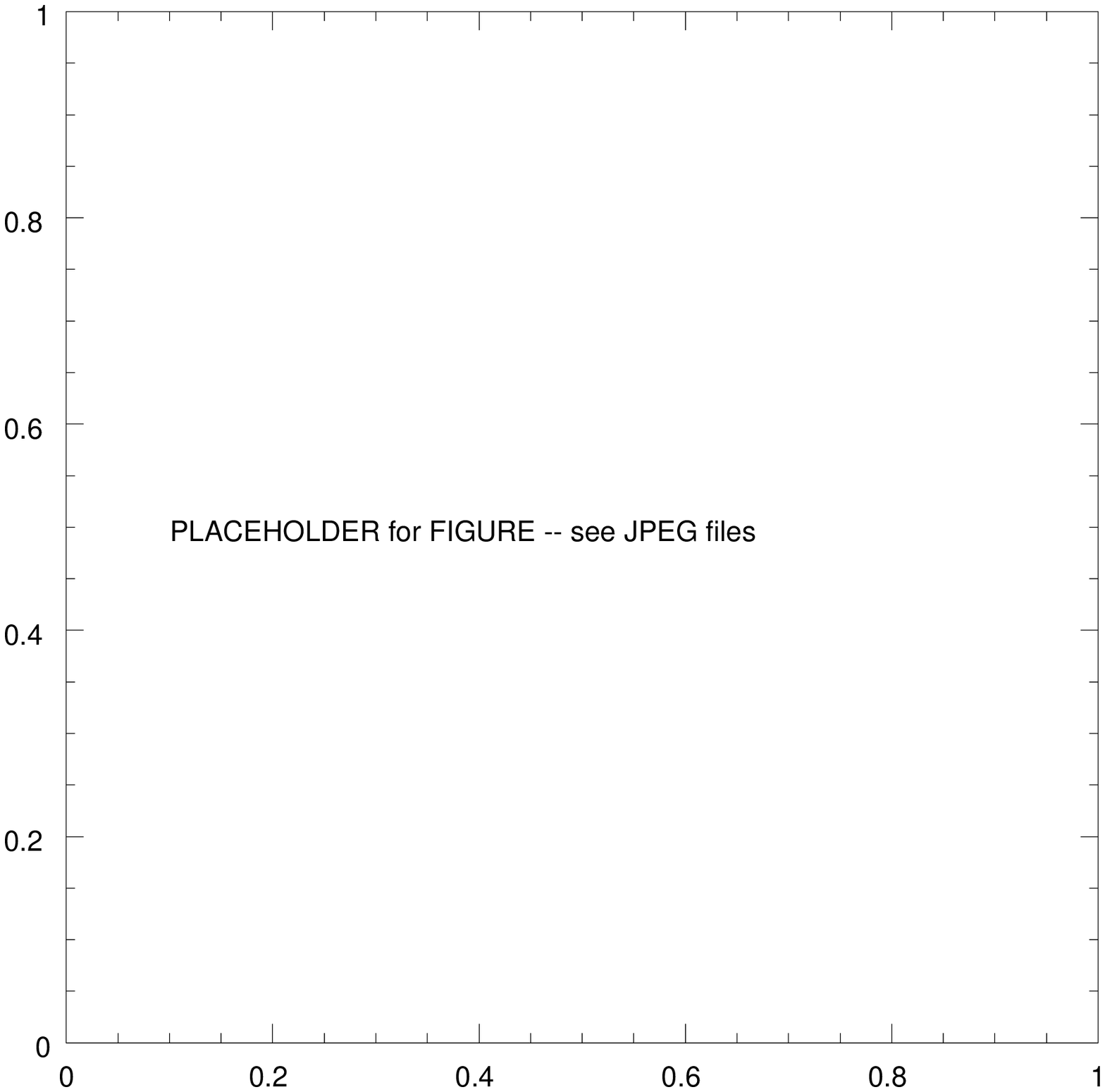,width=18cm}
\caption{Isointensity contours from the adaptively
smoothed XMM-Newton combined MOS image in two broad energy bands on the DSS2-RED plate.}
\label{xmmdss}
\end{figure*}

\section{Results of the data analysis}
The galaxy {IC1262} was observed by  the \chandra\ Observatory
\citep{refchandra} in the
ACIS-S imaging configuration in August 2001 for $\sim$31 ks.  The peak of the
emission was positioned at the aim-point of the back-illuminated 
CCD S3 (CCD7).  A short and unfortunately relatively poor quality observation
was also obtained with XMM-Newton in February 2003
with the Thin Filter for all EPIC
instruments \citep{refxmm, refpn, refmos}.  The log of
the observations is given in Table~\ref{log}.  We base our analysis
on the Chandra observation; however, we will use XMM-Newton data when
needed to confirm/strengthen the Chandra results.  

\subsection{Chandra data}
The Ciao software (Version 3.2.2) was used to clean and update the level 2 event
files and to apply the most recent calibrations, as suggested by the
``Threads" provided online at {\tt asc.harvard.edu}.  
No high background flares were contained in the dataset, so we retain
the total observing time. 
The resulting raw image in the 0.3-7.0 keV  energy range is shown in
Fig.~\ref{rawdata}.
Already in the raw image a complex morphology is evident at the center
of the emission, along with several individual sources.  The iso-intensity contours
of the adaptively smoothed map on the DSS blue image (right panel) indicate a
significantly larger extent than evident in the unsmoothed data. 

\begin{figure*}
\psfig{figure=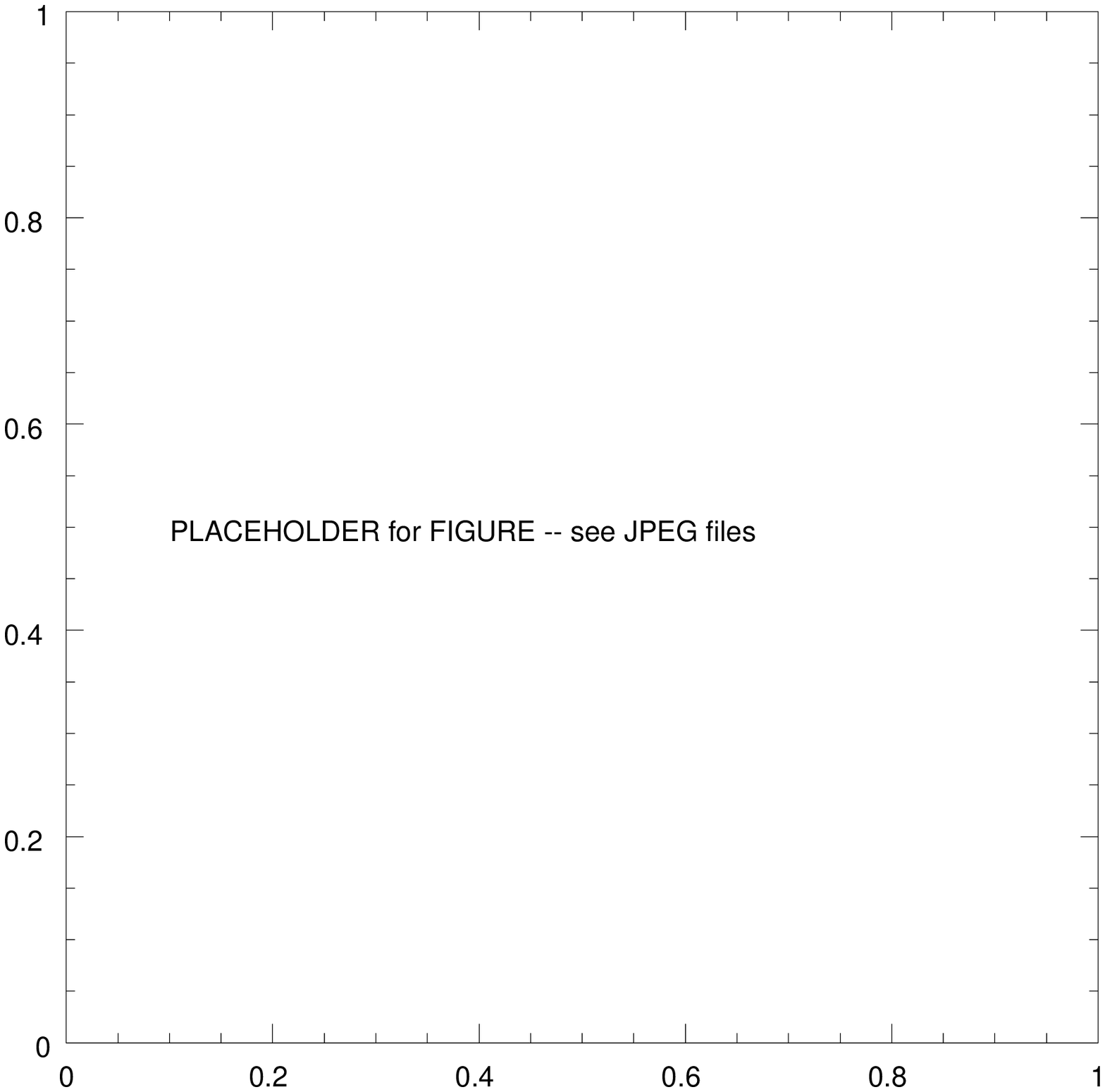,width=18cm}
\caption{Isointensity contours from the adaptively smoothed Chandra 
ACIS-S images in 4 energy
bands superposed onto the DSS2-blue plates.  Top-left: 0.5-0.7 keV;
Top-right: 0.8-1.05 keV; Bottom-left: 1.2-1.5 keV; Bottom-right: 2.0-7.0
keV. 
}
\label{xmaps}
\end{figure*}

\begin{table}
\caption[]{Log of the observations for {IC1262}.}
\label{log}
 
\begin{tabular}{llll}
\hline
\hline
Observ. & Instrument & \multicolumn{2}{c}{Livetime (ksec)}  \\
ID.&& Total & Used \\
2018 &ACIS-S & 30.7& 30.7  \\
 002114091& EPIC-pn & 30.1  & 3.1 \\
& EPIC-MOS1 &44.8 & 3.7 \\
& EPIC-MOS2 &44.8 & 3.4   \\
\hline \hline
\end{tabular}
\\
NOTE: The XMM-Newton times refer to the last observation.  All other
attempts at observing the galaxy could not be used due to
high background.
\end{table}

Since previous X-ray observations of this object with the ROSAT HRI
had already suggested a
large extent \citep[r$>6'$][]{trinchieriandpietsch2000}, 
we expect that a
large portion of the field of view will be covered by the source.  We
have therefore considered blank
sky observations to estimate the field background  \citep[see discussion
in][]{Markevitchetal2003} which we have re-projected to reproduce the orientation of the
{IC1262} observation.  To normalize them to the {IC1262} data we considered
the data in the 10-12 keV range and estimated the count ratios between blank
field and the {IC1262} observations in the same region; the choice of
the high energies should ensure us that the contamination from the
extended emission is negligible, and we in any case excluded a region
around the peak of the emission in CCD7.   We have considered both CCD7,
and the two adjacent front illuminated CCD's, CCD6 to the SW and CCD8 to the 
NE, which should contain some of the extended emission from
{IC1262}, given its extent. 
The rescaling factors that we obtain from the count ratios are remarkably close 
to the values obtained by a simple
rescaling based on the observing times for all three CCDs.   As will
become clear from the comparison of
the radial profiles (see Section 2.3), the shape and normalization of the
count distribution in CCD8 (which we will not use for scientific
analysis)  match to $<10$\% those of its relative normalized background
field; we therefore will assume that the blank fields rescaled by the
counts/time ratios give a good approximation of the shape and level of
the background expected in the field.

\begin{figure*}
\resizebox{12cm}{!}{
\psfig{figure=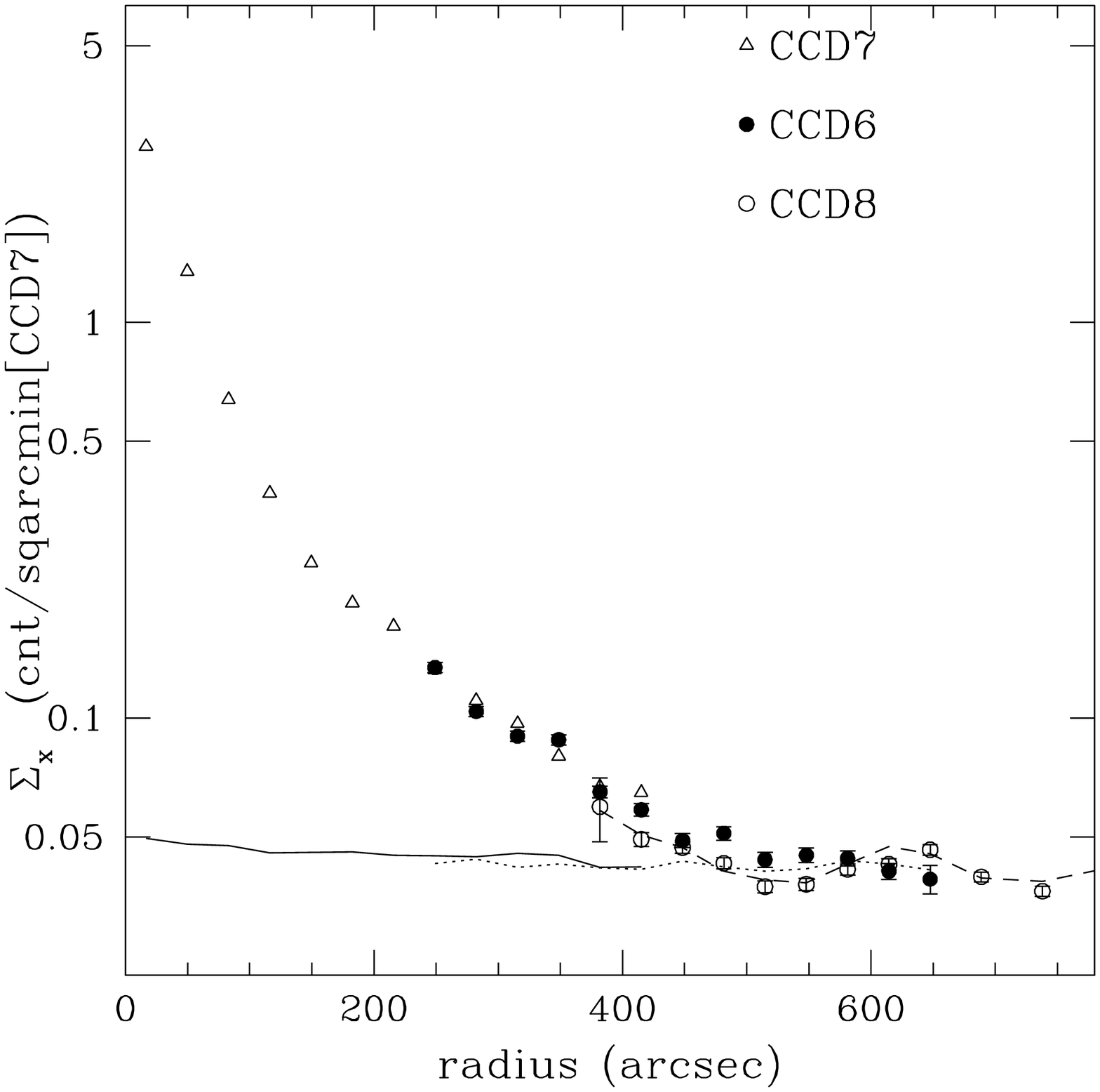,width=6cm}
\psfig{figure=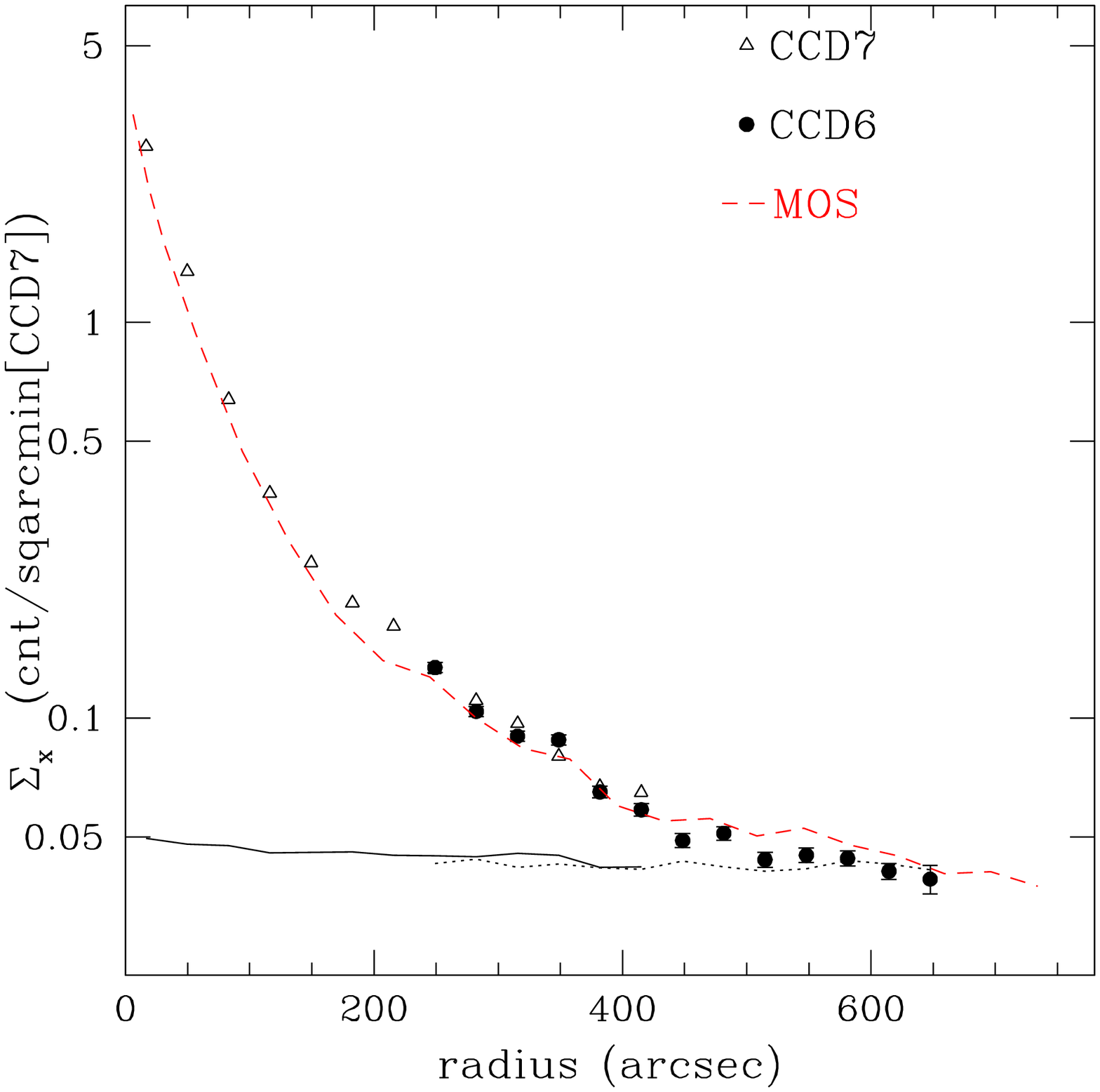,width=6cm}
}
\caption{Azimuthal averaged profiles of the total emission detected in
the field of {IC1262}, in the 0.3-7.0 keV band.  Left: Data are obtained
from the different CCD's in the Chandra ACIS-S detector.  All counts are
normalized to match the values obtained from CCD7 as explained is the text.  The CCD7
background (solid line) is obtained from blank fields, rescaled by the mean of the
ratios between observing times and observed counts at high energies in a
selected region (see text).  Dotted-dashed lines indicate the background levels
for the CCD6/CCD8 fields.   Right: Same as left, without the CCD8 data.
The dashed (red) line shows the profile obtained from the combined
XMM-Newton MOS1+MOS2 data in the same energy band, arbitrarily
normalized to match the Chandra counts. The dotted line indicates the
background for the XMM-Newton observation.}
\label{radtot} 
\end{figure*}

\subsection{XMM-Newton data}
Due to an overwhelming contamination by high background events, we have
used only the last of four attempts at observing this field.
Unfortunately even this observation had to be cleaned from long-lasting 
and strong flaring events. 
We used the standard background light curve data provided
by the SAS software to select the good time intervals where high energy
count rate is stable, although not at the level observed in periods of
real low background (count rates of $\sim  0.8-1.6$ for MOS, instead of
an average value of $\sim 0.8$, and
$\sim 3-12$ cnt s$^{-1}$ for PN, instead of $\sim3$). 
After cleaning, only $\sim 3500$  s of data remain, from the original request
of 20 ks.  Although of limited value on their own,  they confirm and
extend at larger radii the results obtained with the Chandra data, as will 
become apparent in the discussion of the individual issues.

To increase the statistics, we have merged MOS1+MOS2
event files, less affected by 
the CCD-gaps pattern than the PN observation, and we use the result
to produce smoothed maps and radial profiles of the emission,
along with the PN data. 

\begin{figure*}
\resizebox{18cm}{!}{
\psfig{figure=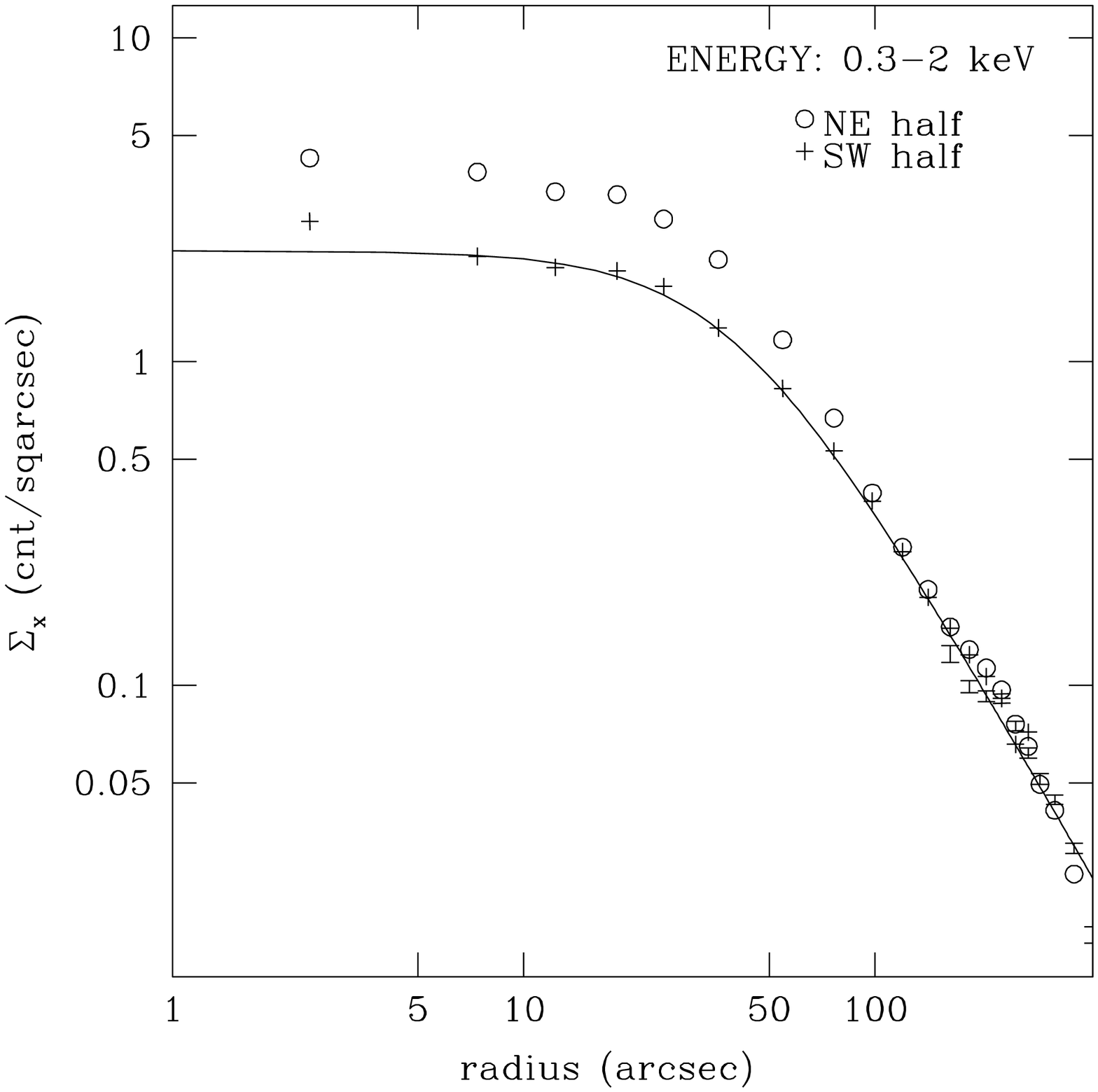,width=9.4cm}
\psfig{figure=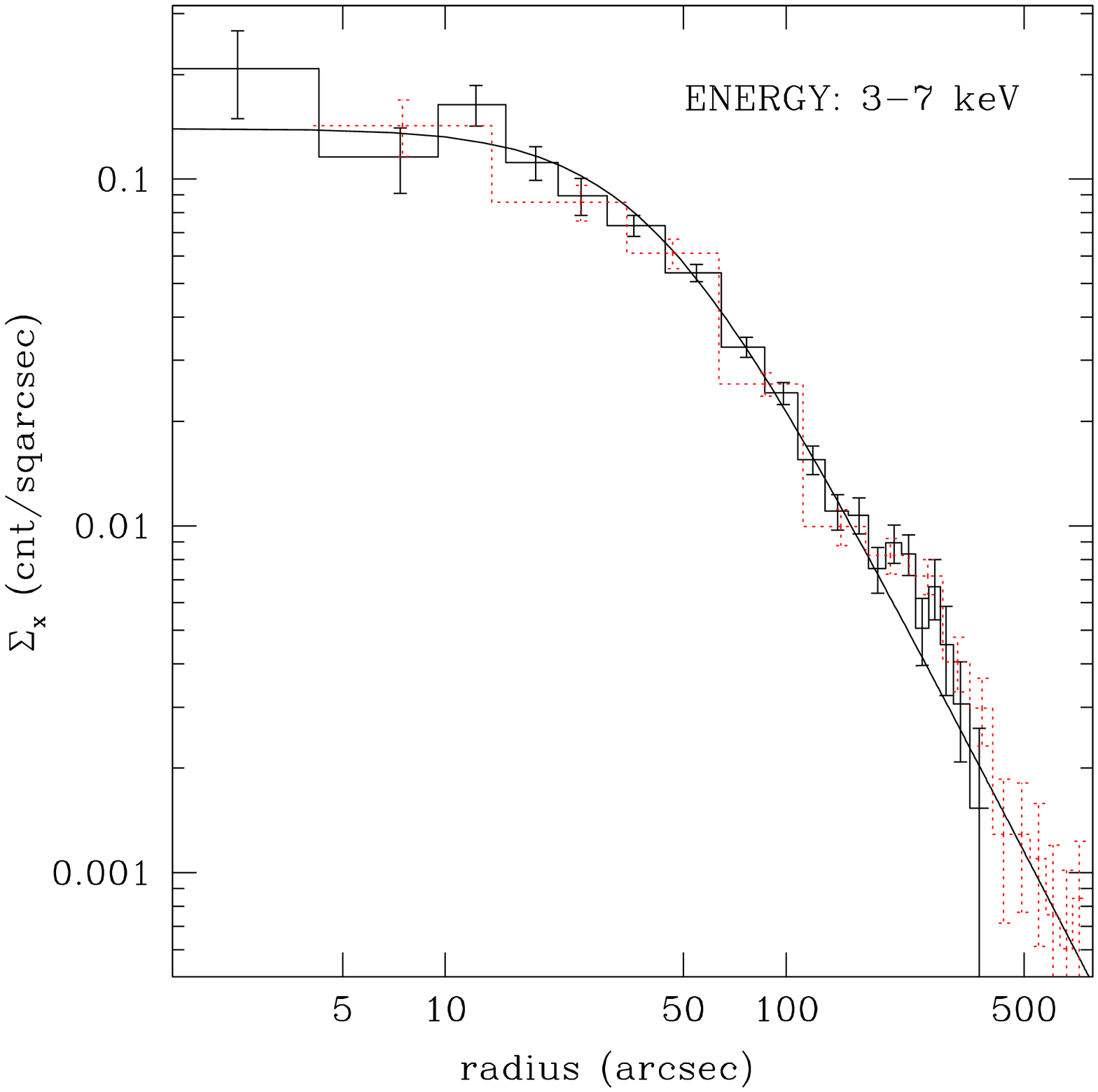,width=9cm}
}
\caption{Net count radial distributions of the Chandra CCD7 counts 
in different energies and different regions as shown
by the labels inside each panel.  The solid line represents a
"$\beta$-profile" with core radius r$_c=40''$ and exponent=0.95
(see text),
normalized to the data. The dotted histogram in the 3-7 keV energy range
is derived from the XMM-Newton MOS image. }
\label{radial}
\end{figure*}

\subsection{X-ray images}

The broad band Chandra image of Fig~\ref{rawdata}-right shows both the
complex central morphology evident in the raw data and the extent of the
emission, that covers the full CCD field of view.  An even better sampling
of the emission at large radii is provided by the bigger field of view of
the XMM-Newton data.  The isophotes of the adaptively smoothed maps in
two broad energy bands shown on the optical image in Fig.~\ref{xmmdss}
immediately suggest that the large scale structure is consistent with
a relatively unperturbed  emission from a group-like potential, in
particular at energies above 2.5 keV.

Adaptively smoothed images of the central region
in different energy bands from Chandra data  are presented in
Fig.~\ref{xmaps}, which clearly illustrates the dependence of the
morphological features of the emission on energy.  
Since the use of adaptive smoothing, an excellent tool for displaying purposes, could be
misleading, and selectively enhance/depress features as a function of
the statistical significance of the data
considered, we have tried to compare images with comparable statistics.  The high energy data is
the strongest limitation, with $\sim 6700$ counts,
so we present a selection of narrow band images at low energies with $\sim 5700 - 10000$
counts each.
We have also used projections of real data in different energy bands, again 
in slices of
comparable statistics, to quantify the reality of the visual impression given by Fig.~\ref{xmaps}. 
In spite of the considerable loss of statistics, that would produce sharper and better defined
shapes, each narrow band image shows a different morphology, which are all different
from the hard band image.  The comparison 
points to a dependence of the central structure with energy, which becomes 
less prominent with increasing energy.    
At energies above 2
keV little remains of the perturbation seen most evidently in the
softer images ($\sim 1$ keV), and the emission is oblong in the E-W direction, relatively well
centered on IC1262.
This is again evidence that at higher energies the emission  is strongly
related to the galaxy and the group potential, and might represent the more
relaxed, ``more traditional" component of a hot
InterStellar/InterGalactic medium commonly found
in E-dominated groups \citep{Mulchaey} down to the inner region around
IC~1262.  

At softer energies the emission around the galaxy is dominated by a complex
structure.  The comparison with the optical image, from the DSS2 plate,
indicates little coincidence between X-ray features and
the galaxies in the group: the coincidence of a few X-ray point sources
with optical compact objects guarantees that the astrometry is good,
and leaves little doubt that the arc-like structure is E and well
outside of {IC1262}.  

\subsection{Source extent}

We compared the radial
profiles of the emission from the source to those of the background in
different CCD's to evaluate the total extent of the source.
Fig.~\ref{radtot} shows the comparison of the azimuthally averaged
profile in the 0.3-7 keV band, both for {IC1262} and the background files.
Different symbols are used to distinguish different CCDs. All counts are
rescaled to match the counts in CCD7.  We have used a 
single point in the profile to normalize them to the CCD7 data: 
CCD6 counts are normalized at r$\sim 380''$,
CCD8 at r$\sim 450''$.  
To ensure that the profiles cover only illuminated parts of the CCD, we have
masked out the regions outside the respective fields of view.  Therefore
the azimuthal average covers 360$\degr$ only out to a radius of $\sim
200''$.  Outside this radius, CCD7 covers a progressively smaller angle
to the NNE, CCD6 to the SSW, and CCD8 an even smaller angle again to the
NNE. 

Given the highly disturbed central morphology, we have used the higher
energy profile, less affected by the central structure, 
to determine the centroid of the emission.  Even with
this choice we are somewhat sensitive to the presence of a possible
source at the center of the emission that slightly biases the
determination of the center of the large scale emission.   Excluding
this potential contaminant, we centered the profile at
17$^h$33$^m$02$^s$.86,43\degr45$'$34\farcs33. This is consistent with the
optical position of {IC1262} given by NED\footnote{\tt
http://nedwww.ipac.caltech.edu/}. 
Other point sources in the field have been
masked out from the profile.

As evident from the figure,  there is very good agreement between CCD7 and
CCD6, at the same radial distance, so we are confident that we can use
CCD6 data to estimate the extent of the emission. 
On the other hand, the radial distribution of the counts in CCD8 shows a modulation 
which is most likely due to residual features related to the flaw in the
serial readout of the chips (see the discussion available on the 
{\tt asc.harvard.edu}
site).  However, the same is true for the background data: in fact the
two distributions look remarkably similar both in shape and in
normalization, so we will no longer use CCD8 data.

\begin{figure}
\psfig{figure=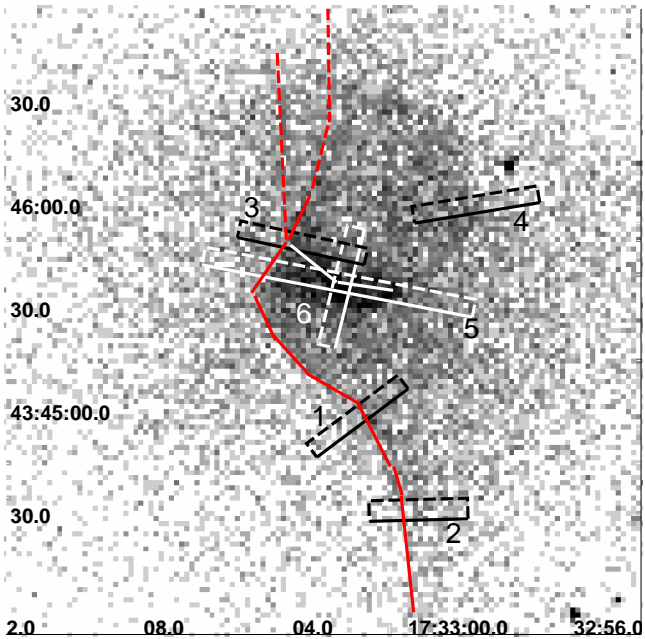,width=8.5cm,clip=}
\caption{Positions of the regions used to derive the profiles shown in
Fig.~\ref{proj} plotted onto the Chandra raw image.  A rough
sketch of the narrow sharp feature at the center of {IC1262} emission
is also given.}
\label{projreg}
\end{figure}

The azimuthally averaged ACIS-S profile indicates an extent out to
$\ge 500''$, after which the shape of source and background profiles
become very much alike.  This is confirmed by the XMM-Newton data
(Fig.~\ref{radtot}-right), that show a radially decreasing profile
with a flattening at about the same radial distance.  We will therefore
assume a total radial extent of $\sim 9'$ for {IC1262},  larger
than reported previously \citep[6$'$,][]{trinchieriandpietsch2000}.
This indicates that the source has an extent of at least 350 kpc,
typical of poor groups.

Figure~\ref{radial} shows the radial distribution of the net emission from
CCD7 in different azimuthal sectors and energy bands. 
The azimuthally averaged high energy profile (right panel) shows that the
emission is well fit by a ``$\beta$-profile":  the line plotted represents  
a r$_c=40''$ and $\beta \sim 0.5$ model; this is not a proper fit to
the data but is simply an example of a reasonably good representation of
the large scale photon distribution.  Again the XMM-Newton data is in
full agreement, and extends the profile at larger radii.  The softer emission,
obtained in two halves, can also be represented by the same law at
large radii: in fact, outside of r$\sim 1\farcm5$, the two halves become
indistinguishable and well represented by the assumed profile.  Inside
this radius, the NE half shows an excess over both the other half and
the parameterization, while the SW direction shows an excess only
within the innermost 5$''$.
The $\beta$-profile values used here are entirely consistent with the
values found in the ROSAT data \citep{trinchieriandpietsch2000}. 

\subsection{Characterization of the X-ray features}

We have considered a more quantitative assessment of two 
parallel aspects of the complexity of the X-ray
emission of {IC1262}: the morphological and spectral 
structures.

\begin{figure*}
\psfig{figure=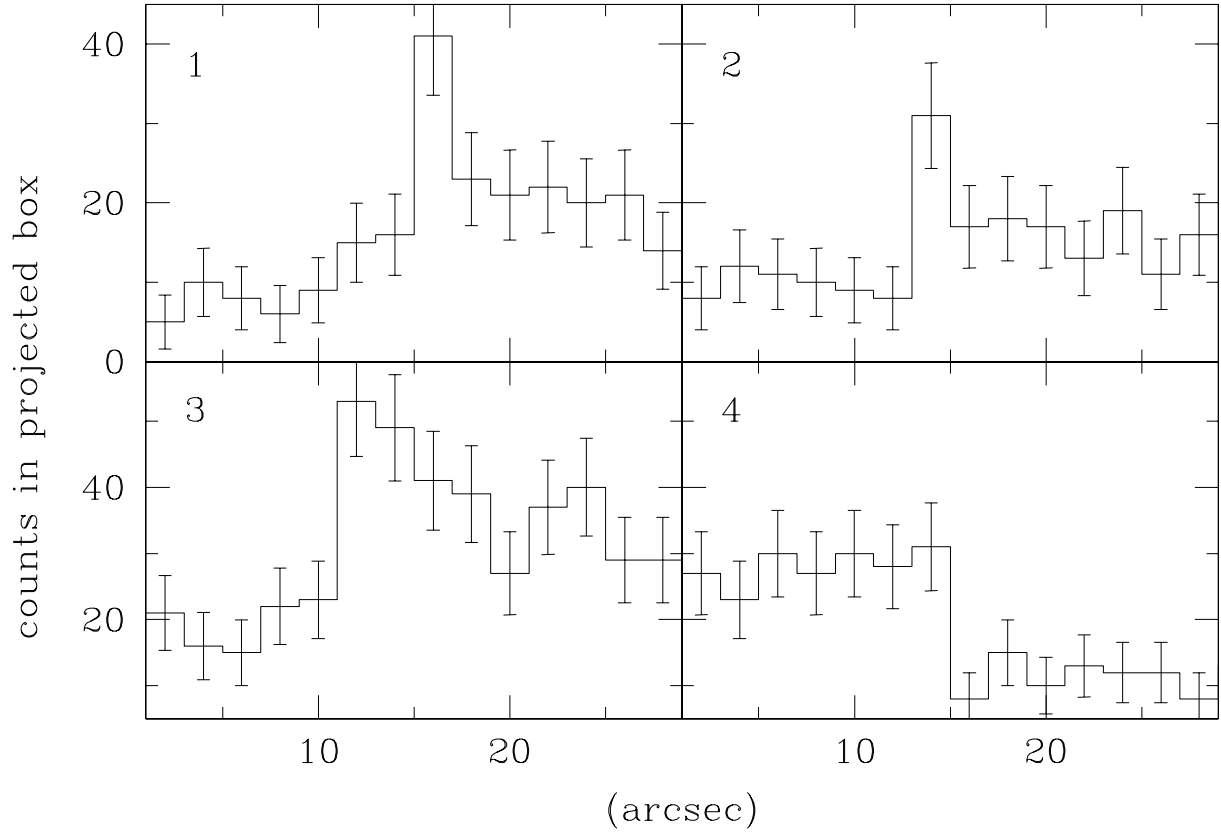,width=17cm,clip=}
\vskip -5.1cm
\caption{Projections obtained in the regions shown in Fig.~\ref{projreg}, that
cross significant surface brightness discontinuities.  Counts
are obtained from the Chandra CCD7 raw image in
the 0.3-2.0 keV band.  The horizontal axis is in arcsec, and goes $\sim$
E to W (low to high numbers).  Each point
is obtained every 2$''$ along the length of the boxes.  Each box is 
5$''$ thick}
\label{proj}
\end{figure*}

\begin{figure*}
\psfig{figure=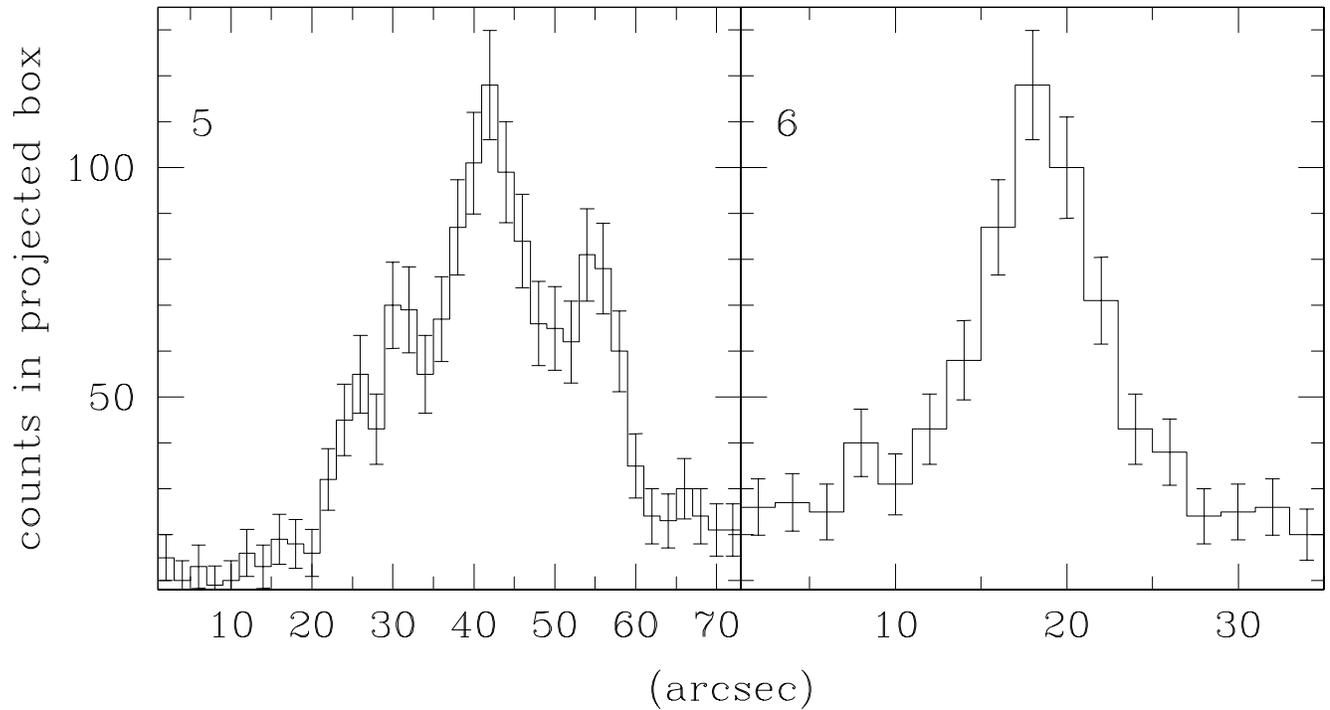,width=18cm,clip=}
\caption{Same as Fig.~\ref{proj} for regions 5 and  6, for which 
we used 6$''$ thick boxes to increase the
statistics.
}
\label{proj2}
\end{figure*}

We have produced a series of  projections across the sharp central
feature along its full length, in regions chosen as perpendicular as possible to the ridge, to
maximize the surface brightness discontinuity.
In Fig.~\ref{proj} we show a small but representative selection, obtained
in the regions shown in
Fig.~\ref{projreg}.  We also obtained projections along and across the
central bright enhancement on and E of IC1262 (Fig.~\ref{proj2}).
Several interesting properties of the structure become apparent:
1) surface brightness jumps of factors of a few in
less than 2$''$ ($\sim 1$ kpc at the distance of {IC1262}) appear to be a
common feature in the data along the whole extent of the structure; 2) the
NS structure can be as narrow as $\le 2''$; 3) there appears to be a
plateau of higher emission ``behind" the jump, 4) the EW enhancement closest
to IC1262 
is approximately 5$\times$20 kpc, with structure along the E-W direction. 

The region of high surface brightness corresponds to significantly cooler
temperatures, as shown by the temperature chart of
Fig.~\ref{colormap} (see following section).  We will discuss these
features in   \S~\ref{discuss}.

\subsection{Spectra}
\label{spectralsection}

\begin{figure*}
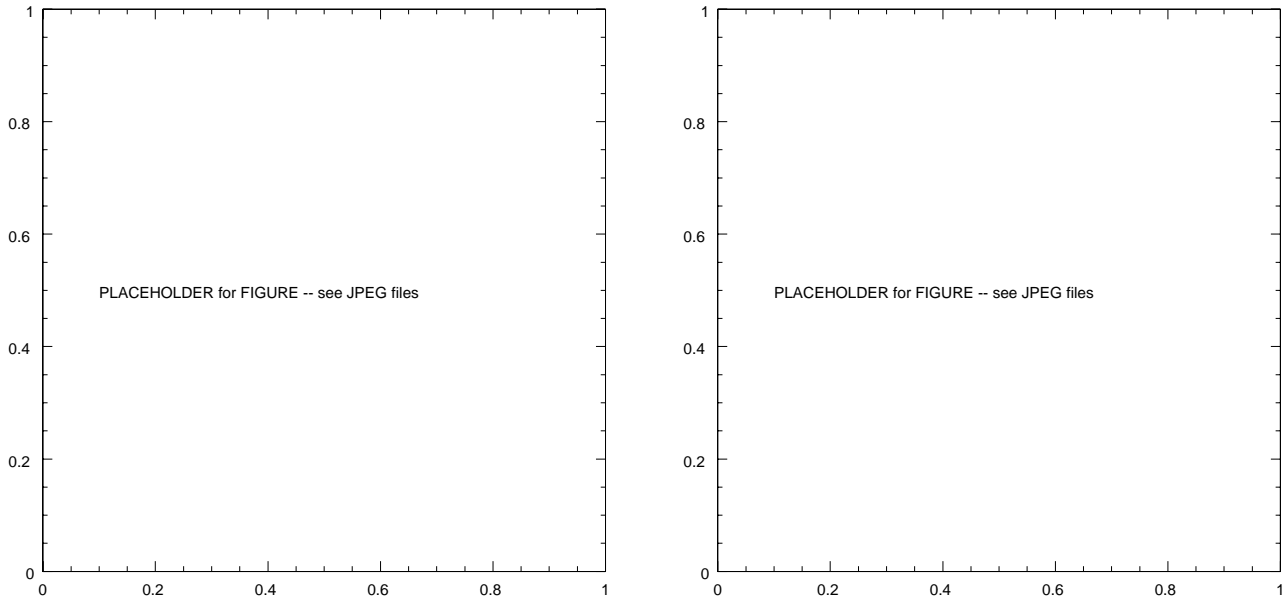

\resizebox{18cm}{!}{
\psfig{figure=placeholder,width=18cm,clip=}
\psfig{figure=placeholder,width=18cm,clip=}
}
\caption{True color images of the inner structure (left) and larger area
(right).  The Chandra CCD7
data have been smoothed with the same function in three
energy bands: 0.5-1.1, 1.1-2.5, 2.5-5.0 keV, represented as red, green
and blue respectively.  
}
\label{colormap}
\end{figure*}

Since the emission detected is significantly higher than the expected
``sky" and detector 
backgrounds, we consider as background the emission around
each region to highlight differences with the neighbouring emission.  
In particular, for all regions within $1\farcm5$ from the
center (see Fig.~\ref{colormap}) we use a circle to the SE, all within
the area delimited by the innermost $\sim 1\farcm5$ circle, 
that does not contain any of the
morphological high surface brightness 
features (included in the many smaller regions in the figure).
We have checked that its energy distribution is consistent with that of
other similar regions, chosen always within region ``inner" but away from the
features evident; the choice of the particular location is simply dictated
by the fact that to the SE we can choose a larger circle 
than elsewhere.
The outer regions have been used for the inner larger annuli, so region
``middle" is the background for region ``inner", and region ``outer" for
``middle".  A separate region to the NE is used for the regions further
out (``outer",``N",  ``E" and ``NE").

\begin{figure*}
\resizebox{18cm}{!}{
\psfig{figure=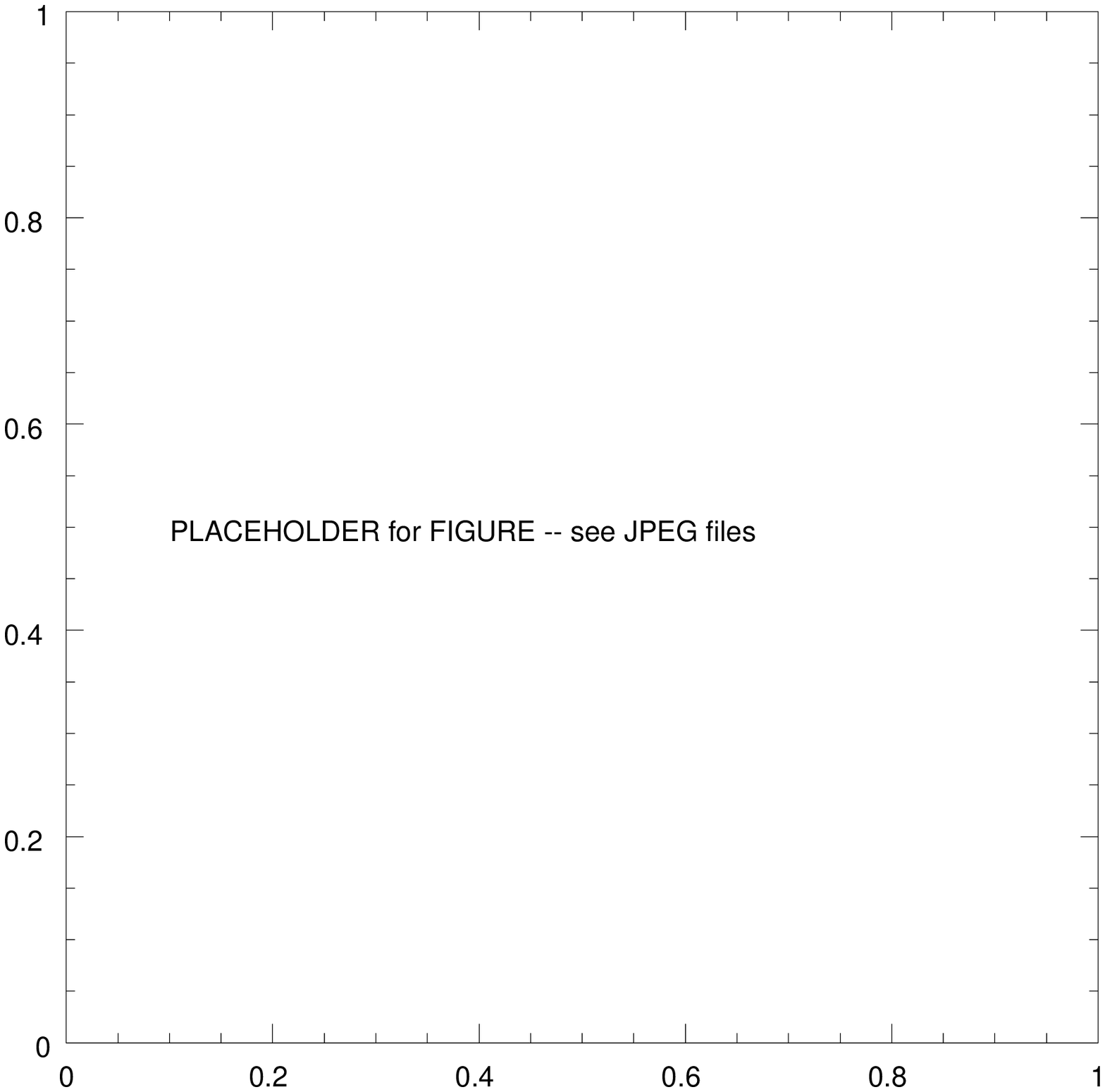,width=\columnwidth,clip=}
}
\caption{Schematic results of the temperature
distribution in different regions, for the innermost (left) and outer
(right) regions considered. The values indicate the 90\% confidence
interval of the best fit temperature parameter in keV.
The straight line at the bottom indicates
the edge of the Chandra CCD7 and the end of the spectral region considered.}
\label{temps}
\end{figure*}

We have defined several regions of emission 
from which we derive both
the contribution of each individual component 
and the  photon energy distribution 
to be used for spectral analysis, using the
color map shown in Fig.~\ref{colormap} as a guide. 
Figure~\ref{colormap}
is obtained from the combination of three images in 0.3-1.1,
1.1-2.5, 2.5-4.0 keV bands.  The data are adaptively smoothed with the
same smoothing function to reduce the noise.  Although colors might not be
easy to translate into temperatures, the map clearly indicates that
the central region  is extremely complex, and that
even outside the centre there is structure in the
``colors". 
The regions defined and the temperatures we derive for each of them are
summarized in Fig.~\ref{temps}. 
Photons have been assigned only to one region 
so that each ``external" region
does not contain the inner ones.  Point-like sources evident in the maps are also excluded.

We have binned the raw data with the dual purpose of  minimizing the
bin-to-bin fluctuations (using only a few channels, even when the
statistical significance of each would be sufficient for the correct
use of the chi-square test, mostly at low energies) and of  obtaining a
positive signal in all spectral channels considered, after  the background
is subtracted (larger bins, usually required at high energies, where
the background contribution is most relevant).

In  most cases the data can be fitted with a
single temperature plasma.  When the statistics is good we have let
all parameters free to vary (low energy absorption, temperature and abundances).
In all cases absorption is consistent with galactic N$_H$.  Best fit
abundances also span a rather narrow range around 0.3-0.5 $\times$solar. 
When the data quality is poorer, we fixed N$_H$ at the galactic
value and the abundance parameter at  0.5 (and checked also the results
for 0.3 and 1) $\times$solar.  We have used the \citet{Wilms} tables for
all fits. 

The  90\% confidence values for the temperatures derived for each region,
shown graphically in Fig.~\ref{temps},
indicate that: 1) the central feature has a
systematically lower temperature, around kT$\sim$1 keV,
than the surrounding regions, even
close-by ones at lower surface brightness; 2) there is
a trend of increasing temperatures with radius, from an average 1.7
keV in the central region to 2 keV outside the inner $1\farcm5$; 3)
cooler temperatures can still be traced out to $2'-3'$ 
along what appears as the extensions of the tail to the S and N of the
feature, in spite of little evidence in the morphology at those radii
(but some distortions are visible in the isophotes in Fig.~\ref{rawdata}). 

These results are in broad agreement with those published by
\citet{Hudson},  who find that most of the regions sampled
require a two component model. This is entirely consistent with the
presence of a softer high density feature on top of a broader harder
emission pictured in Fig.~\ref{temps}. However they suggest that a
better representation of the data, and more physically meaningful, is
found by adding a non-thermal component in the region  to the north and
south of the X-ray flux peak, rather  than a second thermal component, to
account for excess emission above a $\sim 1.3-1.4$ keV plasma emission.
Since we have used a much finer sampling of the area, with smaller size
regions, and have identified the softer component with the region of the
perturbation, we identify the harder one with the large scale diffuse
emission and therefore require a second thermal model to fit the data.  
Moreover, we have checked that the  $\sim 2 $ keV  model is a good fit at 
different azimuthal angles (we have done it for region ``middle", and
for the innermost circle, as explained above), and not just in
one privileged direction, which supports the interpretation with emission from a
pervasive hot gas.

\section{Discussion}
\label{discuss}

\begin{figure}
\psfig{figure=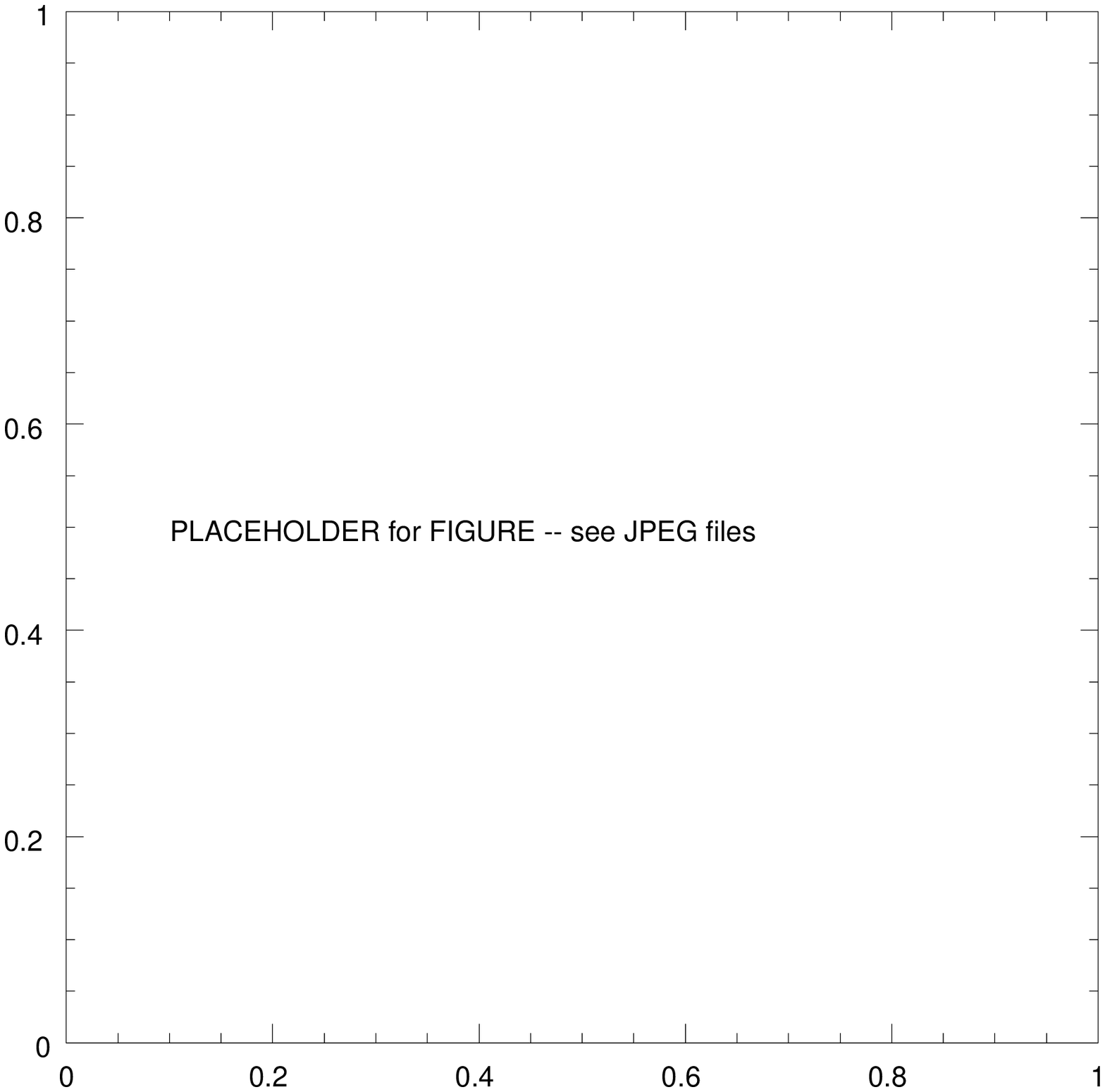,width=8.5cm,clip=}
\caption{
Spatial distribution of all galaxies with measured redshifts in the field.
Small labeled circles (red) indicate background galaxies ($>18000$
 km/s).  Squares (magenta) and circles (black) are members with v $<$ 9500
 and v $>$ 9500 km/s
 respectively. IC 1262, IC 1263 and IC 1264 are indicated.
}
\label{velocities}
\end{figure}

{IC1262} is the dominant early type galaxy at the center
of an aggregate of galaxies with 31 confirmed members \citep{smithetal}
distributed over approximately 20$'$ (see Fig.~\ref{velocities}). While 
a reliable census of membership is incomplete \citep[for details, see
the selection criteria in][]{smithetal}, 
it is clear that IC1262 lies at the 
center of a rich group or poor cluster, with only a few 
galaxies of comparable magnitude (e.g. IC1263 $4'$ to the
NE, and  IC1264,  $8'$ to the S, using the photometry given in NED).  
The group has 
a regular velocity
distribution \citep[cf. also Fig. 7 in][]{smithetal} and an estimated velocity
dispersion $\sigma$$\sim$560 km s$^{-1}$ \citep[S. Andreon, private
communication, using][]{beersetal90, andreon}.

The total X-ray luminosity of the group, as measured from the ROSAT data
\citep{trinchieriandpietsch2000} and
confirmed by the Chandra and XMM-Newton data, exceeds L$_X$=$10^{43}$ erg
s$^{-1}$.  With a gas temperature kT$> 1.6$ keV, the group  falls at the upper
end of  the $\rm L_x-T_x$ and L$_X$-$\sigma_V$ 
relations determined for Group Evolution Multiwavelength Study (GEMS)
groups and close to the values for richer clusters
\citep[$e.g.$][]{OsmondPonman2004}.

The spatial distribution of the extended gas, measured at energies above
2-3 keV, can be modeled with a $\beta$-profile.  The core radius
we have assumed in Fig.~\ref{radial}, r$_c \sim$25 kpc, is consistent
with values obtained for the GEMS groups for the   group-like component
\citep{OsmondPonman2004}.
Many groups
could also display a central  component, consistent with a contribution
from the  
central galaxy, with a  $\sim 10 \times$ smaller
core radius and located  at the center of the more extended one 
\citep[e.g.,][]{Mulchaeyzabludoff98,HelsdonPonman2000}.
The complexity of the X-ray structure near IC1262 makes it difficult
even with Chandra resolution to map the central part of 
the diffuse gas component. Above 3 keV, where the central complexity is
almost absent, we find no indication of a ``shoulder" in the profile that is
usually the signature of the central galaxy. However, such a signature might not
be visible above keV if such a component has a softer spectrum as might be
expected from the InterStellarMedium (ISM) of an early-type galaxy.
The XMM-Newton data, both at high and low energies, confirm the Chandra results and 
extend the profile at least out to $\sim$ 350 kpc (Fig.~\ref{radial}).

The morphology of the extended gas in the inner $1'-2'$ is
surprisingly complex with a narrow, sharp and cool structure.   This is
most likely a signature of recent significant  events
imprinted on an  underlying relaxed component.  
The agreement between the centroid of the hard energy
emission and IC1262 supports the idea that the overall system is in
equilibrium. In that case the central structure is a perturbation
with several interesting characteristics: 1) 
surface brightness ``jumps" of factors of $>2$ and  2) sharp edges
(see Fig.~\ref{proj} and \ref{proj2}); 3) long coherent 
structures in the form of a) a $\sim 2''-4''$ ($\sim$ 2 kpc) narrow ridge, that
can be visually traced
for $>2\farcm5$ ($> 200$ kpc), as roughly sketched in
Fig.~\ref{projreg};
an extension to this could also be found $\sim 2'- 3'$ N, where a region
of lower temperature is visible (see Fig.~\ref{colormap}), and b) 
an almost complete ``loop" or an arc on and N of IC1262, with a
radius r$\sim 30''$ (20 kpc); and 4) 
cooler temperatures along the full length of the
high surface brightness ridge, relative to the average value in the
lower surface brightness unperturbed inner region (see
Fig.~\ref{temps}). 

We can use the spectral results 
to estimate the gas density in the perturbation, which we consider on
top of the
unperturbed component.  
 We have used the volume corresponding to the region shapes
shown in Fig.~\ref{temps}.  Since the third dimension is unknown, we assume an extension 
along the line of sight similar to that in the transverse direction.  
We find n$_e \sim
0.02-0.03$ cm$^{-3}$ in the arc/loop, and values in the range n$_e \sim
0.015$ close to the center to n$_e\sim 0.003$ cm$^{-3}$ at the south tip 
along the NS ridge.
The total 
luminosity of the perturbation is  L$_x \sim 1.5 \times
10^{42}$ \ergsec\ (0.5-2.0 keV), 
excluding the
contribution from the point sources embedded in them, and could involve
a mass M$_{gas} > 6 \times 10^9$ M$_{\sun}$.   Assuming radiative cooling
as the dominant energy loss mechanism, the structure could have a  cooling
time $\tau \sim
10^9$ yrs. 
Although not
measured, some degree of concentration is consistent with the appearance
of the emission, and would result in higher densities and lower gas mass
involved. 

There is a general trend of higher pressures along and across the
perturbation region than in the surrounding hot ICM gas, mainly due
to somewhat higher densities along the X-ray ridge. Here the derived
pressures range from $2\times 10^{-11} \, {\rm dyne} \, {\rm cm}^{-2}$
to $1 \times 10^{-10} \, {\rm dyne} \, {\rm cm}^{-2}$, while the
average pressure in the inner region or just outside it 
is typically at $\sim 1-4 \times 10^{-11} \,
{\rm dyne} \, {\rm cm}^{-2}$.  The overpressure in the high density
regions  will reduce due to expansion.  For a temperature of 1 keV,
the isothermal speed of sound will be roughly 400 km $\rm s^{-1}$, and hence
an expansion of about 0.4 kpc per Myr can be expected, decreasing
strongly as the pressure difference with respect to the ICM becomes
smaller. The ``sharpness'' of the feature can be maintained easily in
regions with a small pressure difference or by an additional pressure
component due to ICM turbulence or due to a magnetic field.

The timescale for the dissipation of discontinuities may cover
a wide range, depending on the processes at work and the kind
of discontinuities under consideration. If we have a tangential
discontinuity, the stabilizing effect of a magnetic field would inhibit
mixing due to magnetic tension, depending on the magnitude of the field.
This could be one reason for the observed ``sharpness''
of the NS feature.  If on the other hand the gas in the feature reaches a
similar pressure than its surroundings, diffusion of the contact discontinuity
progresses slowly and may be somewhat enhanced by a fluid instability
(e.g.\ Kelvin-Helmholtz, bending mode).  The diffusion coefficient
is typically of the order of the diffusion speed (e.g. turbulent
velocity) times the gradient scale (e.g. density or temperature).
The reason that the feature would still remain sharp could then
also be attributed to the fact that the density near the edge of the
diffusion boundary drops rapidly, and the X-ray emissivity decreases
quickly to the background level.

What can the origin of this feature be?   We have looked for analogies
with other better studied systems, where information in several
additional bands
(e.g. H$\alpha$, dust emission, radio structures) is available, to 
understand its nature.  Indeed we found increasing evidence of
structures in the IGM of groups and clusters of
galaxies, that bear  resemblances to this one, but also show 
clear differences. 
We examine here some of the most relevant phenomena that could explain
the observational evidence.

\subsection{{Shock}:}  Since clusters of galaxies are thought to grow
through gravitational infalls and mergers of smaller groups, shocks are 
expected to be common in groups and clusters.  Indeed they are now imaged in the
X rays: the Chandra image of 1E0657-56
\citep{Markevitchetal2002}
shows a clear cut example of a bow shock; the optical, IR and
radio evidence of the shock in Stephan's Quintet (SQ) is now very well
documented in the X-ray images as well
\citep{trinchierietal2003,trinchierietal2005}.

While the morphology of the structure could be indicative of a shock, 
its temperature, significantly cooler than
the ambient medium, is a problem.  In this respect, it could be analogous
to the shock in SQ, which is also significantly cooler than predicted
from the impact 
velocity of the galaxy that produces it.   As extensively discussed in
\citet{trinchierietal2005} the discrepancy cannot be easily overcome,
unless we can resort to additional and very efficient cooling or peculiar
geometrical effects. 
Most of the arguments used for SQ can be applied here as well, and again 
they are hard to reconcile with the shock
interpretation. 

An additional concern comes from the dynamics of the system.  
The distribution of the member velocities is regular \citep{smithetal},
and although
velocity differences of the order of 1000 km s$^{-1}$ between members are
observed, they are consistent with the velocity  distribution from a relaxed
and massive group. 
Neither the optical distribution of galaxies around {IC1262} nor the
distribution of redshifts would indicate peculiar motions in the system
as a whole.  
In particular Fig.~\ref{velocities} shows that, even when we arbitrarily 
separate high and low velocity objects, we do not see 
any systematic effect that
would suggest collisions or interactions between two systems.   Unlike the
SQ system, there is no evidence of a high velocity sub-system that could be 
slamming into the hot gas and produce the shock,  
unless the motion is all in the plane of the sky.  Even in this case,
there is no ``obvious" structure which can be called in question:  as
already pointed out, the nearly NS distribution 
of objects, which could suggest that they all lie along the 
perturbation, could be a result
of the selection, and not reflect the true distribution of member galaxies
in the {IC1262} system.  A deeper image of the field, with proper
identification of all potential members, is badly needed to clarify this. 
The lack of optical evidence of peculiar motions,
or of high velocity ``intruders" or large scale shocks that were
instead unquestionable in Stephan's Quintet, is a serious concern for
such an interpretation.

\subsection{{Cold front/filament}:}
There is increasing evidence in the literature of sharp discontinuities
in the surface brightness of the diffuse hot IGM  component in clusters
of galaxies. These have lower temperature than the surrounding
medium, and have therefore
been named ``cold fronts"  
(Vikhlinin et al. 2001, Mazzotta et al. 2002). 
The origin of these features is also  attributed to merging of smaller
structures with 
larger systems, mostly in the plane of the sky,  or even merging between major
systems \citep{mathisetal, bialek}.  The fronts have been
interpreted as boundaries of dense cores moving through a hotter ambient medium
\citep{Markevitchetal2000,  Vikhlininetal2001}.  They are very sharp
features, with lower temperatures on the brighter side of the
discontinuity. Ram pressure stripping is thought to play a role also in
giving the structure the characteristic curved shape observed
\citep{Markevitchetal2002, Heinzetal2003}. Merger cold
fronts have been reproduced in recent hydrodynamic simulations
\citep[e.g.,][]{bialek,nagai2003,mathisetal}.

Cold narrow filaments have also been observed in a few cases: while
some are associated 
with optical line emission (H$\alpha$) as in 
in A1795, \citep{crawford}, and in Perseus \citep{fabianetal2006},
a narrow arc-like feature is observed in A3667 \citep{Mazzottaetal}
and interpreted as evidence of large scale hydrodynamic instability
developing from a prominent ``cold front" due to a large  cool cloud
moving in the cluster atmosphere.

In {IC1262} the morphology is more reminiscent of a cold narrow filament,
since most ``cold fronts" appear as more compact structures with a very
sharp ``front" and a more gradual trail-off behind it. 
However, unlike the above examples, the filamentary structure is not directly
connected to optical objects or structures. 
The bright feature EW of IC1262 could be
associated with the galaxy itself, and could be the result of stripping of
the ISM as the galaxy moves into the hot IGM.  However, its location at
the center of the large scale distribution of the gas and its optical
classification as a cD argue against significant motion relative to the
cluster potential, unlike for example the tail observed in the
late-type galaxy \object{ESO 137-001} in \object{A3627} \citep{sunetal}. 
Moreover, the much larger structure could not be interpreted
with stripping of {IC1262}.  We cannot comment on an association with
optical line emission  or with dust, for lack of relevant observations.  
In addition, as explained above, 
we have no direct evidence from the velocity field of the relative motion
of two distinct subgroups that would justify motion of cooler gas in the hot
atmosphere, of the kind expected in cold fronts.   

\subsection {{Current or past radio/AGN activity - cavities}:}

\begin{figure}
\psfig{figure=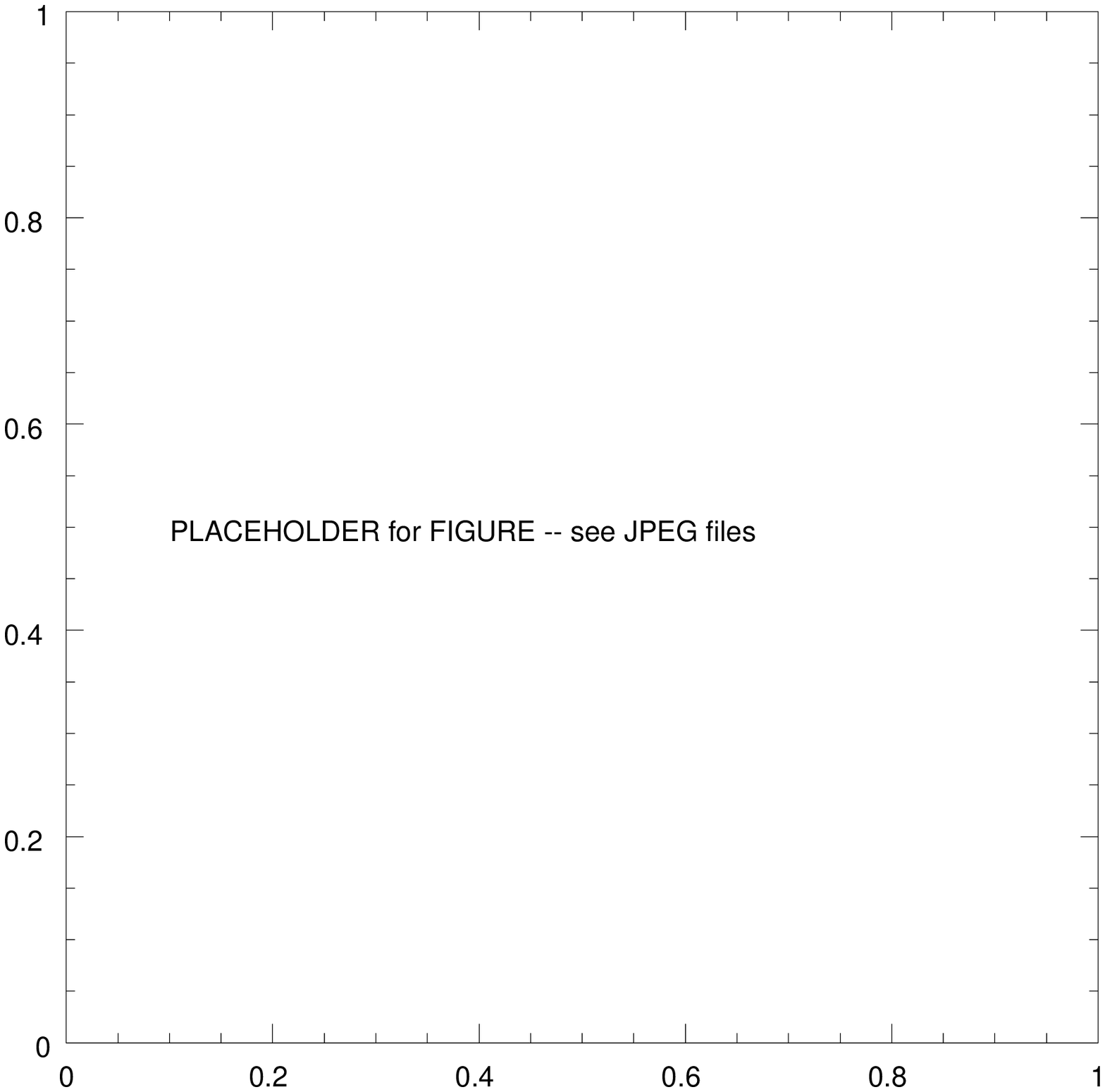,width=\columnwidth,clip=}
\caption{Radio contours from the WENSS 375 MHz map onto the Chandra smoothed image. 
The positions of the 1.4 GHz sources (FIRST, white crosses; 
NVSS, black diamonds) and of the CCD boundaries  are also
indicated. } 
\label{radio}
\end{figure}

Structures associated with powerful radio sources have now been
identified in many clusters of galaxies:  in the Perseus
cluster \citep{Fabianetal2000}, in Hydra-A \citep{McNamaraetal2000},
and in several Abell clusters, A4059 \citep{HuangSarazin1998}, A2597
\citep{McNamara2001}, A2052 \citep{Blanton}, and
A2634 \citep{Schindler1997}, to cite a few.   One of the most
spectacular sources,
Cygnus A, also shows a very complex central X-ray morphology and associated temperature
structure \citep{Wilson}. 
The radio sources appear to act
on the surrounding hot medium by pushing aside the gas and leaving low
surface brightness ``cavities" delimited by bright rims.  In most of the
cases studied, the emission from the rims is softer
than the surrounding gas,  consistent with models in which cooler
material is  displaced as the radio source expands subsonically.
Models that would suggest these structures 
are produced by strong shocks \citep[e.g.][]{Heinz} are instead not
supported by the observations.

The X-ray morphology and spectral characteristics of this system
would strongly suggest that we are witnessing the same phenomenon
here.  A radio structure associated with {IC1262} is visible
in the Faint Images of the Radio Sky at Twenty-centimeters survey \citep[FIRST][]{first}, 
NRAO VLA Sky Survey
\citep[NVSS][]{Condon}, and Westerbork Northern Sky Survey \citep[WENSS][]{wenss}
archival data (see Fig.~\ref{radio}).  At 327 MHz, the low resolution WENSS
data indicate a rather extended source, with a morphology reminiscent
of a radio galaxy centered on  {IC1262} and two lobes north and
south of the galaxy.  The NVSS data are not as extended, but show a
source coincident with {IC1262}, and  two sources N and S,  at
the position of the WENSS ``lobes", with the Southern source indicating
possible substructure.  At even higher resolution, FIRST detects compact
sources at the core and the peak of the Southern lobe (as indicated
in Fig.~\ref{radio}).

The interpretation of the radio structure is not clear, due to the limited
data quality: the two sources N and S have relatively steep spectra
($\alpha_r \sim 1.3-1.6$), while the central source is only visible at
1.4 GHz, indicative of a flatter spectrum.  The comparison between the fluxes
detected in the FIRST and NVSS surveys moreover suggests that both the
central and southern sources are extended.  Whether the radio emission
is due to a double lobed radio source or whether a more complex structure
should be considered
 \citep[which would include a mini-halo near {IC1262},
see][]{Hudsonetal} will require higher resolution deeper observations.
In any case, the luminosity of the radio source, if it is  associated
with {IC1262},  is quite low, P$_{1400~MHz}= 3.8\times 10^{22}$
W Hz$^{-1}$ (Central source only) or P$_{1400~MHz} =2\times 10^{23}$
W Hz$^{-1}$ (including all three NVSS sources), significantly
fainter than the power associated with most of the other radio
galaxies responsible for cavities and rims. Moreover, the length of the X-ray structures
associated with the radio sources is in most cases of the order of tens
of kiloparsecs, rather than the a few hundred kpc size  discussed here, except
possibly for the Cygnus A source, which is several orders of magnitude brighter
both in X rays and in radio.

In a few clusters, the interaction of the IntraClusterMedium (ICM) with a 
radio galaxy becomes less clear: X-ray cavities that do not coincide
with the active radio lobes are found for example 
in Abell 2597, NGC 4636 and Abell 4059
\citep{McNamara2001, Osullivanetal2005, Heinzetal2002} and the outer
regions of Perseus A \citep{Fabianetal2000}.  The presence of these
cavities is now believed to be associated with past, currently faded, 
radio activity, related to previous cycles of
nuclear activity. 

The analogy with these other systems is again intriguing but not entirely
satisfactory: 
Abell 4059, although host of ghost cavities, is currently classified as a 
FRI radio source, at $>1$ Jy, which is considerably brighter than 
{IC1262}, also considering its larger distance. 
\citet{Heinzetal2002} argue that the source has
faded significantly, from a FRII source responsible for the IGM
structure.  On a smaller scale, the structure in  NGC 4636  has also been
interpreted as due to past activity \citep{jonesetal, sunetal}.  The
``current" radio source is comparable in strength with the source in
{IC1262} \citep{BirkinshawDavies, StangerWarwick}.  The major difference
is again in the size of the structures: \citet{Osullivanetal2005}  suggest
a scale of $\sim 20-30$ kpc for the expansion of the cool gas in the
IGM of NGC 4636, and measure a size of $\sim 15$ kpc across the plume
extending southwest from the galaxy core.  The structure in {IC1262}
is $\sim 10\times$ in size!

However, the connection between the radio and X-ray structures is rather intriguing
and worth of further studies. 

\subsection{Ram pressure stripping of a spiral member}

In an exploratory study of parameter space, \citet{st99}
modeled the interaction of a spherical galaxy with the ICM 
and showed that, although stripping is most efficient in rich clusters,
where galaxies might lose all of their ISM due to higher ram pressure
conditions, the best observable cases in X-rays should be cool clusters
or groups of galaxies.  This can be understood as follows: the best
observable signatures should be the bow shock ahead of the stripped
galaxy and the tail of stripped galaxy material.  Whereas the bow shock
is strongest in high velocity dispersion clusters, the actual contrast
in X-ray surface brightness is small due to a more luminous ICM gas.
On the other hand, the tail is strongest in a galaxy moving through a
cool cluster environment, because here stripping is not complete and the
stripped material does not disperse as fast as in the high velocity
case.
In addition, the effect of a high contrast X-ray surface brightness
tail is enhanced by gravitational focusing of the slow moving material
trailing behind the galaxy.

The observational evidence of the long narrow  coherent structure and of
distorted isointensity contours and cooler temperatures in the direction
of the  spiral galaxy \object{IC1263}, located $\sim 4'$  N of IC1262
(see Fig.~\ref{rawdata}),
prompted us to consider it as a 
suitable observable candidate for ram pressure stripping in a cool cluster. 
Several arguments are in favour of this scenario: 

\noindent $\bullet$ the amount of cooler X-ray gas is
comparable to the ISM mass in {IC1263}.  Not much is known about
this galaxy, except that it is of type SBab and that it has
a semimajor axis of $1.7$ arcmin diameter, translating into a physical
size of 68 kpc. 
The HI content of this spiral is not
measured: the galaxy was included in a 21-cm line survey
\citep{theureau}, but no information is given on this object.
We therefore assume
that it has features of a large spiral with a baryonic mass of $\sim 2
\times 10^{11} \,  M_{\sun}$, and $\sim 2 \times 10^{10}
\,  M_{\sun}$ of HI gas, comfortably larger than the masses derived for
the X-ray gas 
(cf.\ Sect.~3). 

\noindent $\bullet$ The conditions for stripping a large amount of gas are favourable.
Using the \cite{gg72} criterion supplemented by a relation for 
the stellar surface density with rotational velocity, $v_{\rm rot}$
\citep{Binney}, we obtain 
\citep[see][]{vo01} for the ram pressure, $P_{\rm ram}$:
\begin{equation}
P_{\rm ram} = \rho_{\rm ICM} \, v_{\rm gal}^2 =
\Sigma_{\rm gas} \frac{v_{\rm rot}^2}{R_{\rm strip}} \,,
\end{equation}  
where $\rho_{\rm ICM}$, $v_{\rm gal}$, $\Sigma_{\rm gas}$ and 
$R_{\rm strip}$ are the mean ICM density, the velocity of the
galaxy relative to the ICM, the surface density of the galaxy's ISM,
and the minimum stripping radius 
of the galaxy (all material at $r < R_{\rm strip}$ will be bound to the 
galaxy), respectively. Using observed quantities such as 
$\rho_{\rm ICM}= 8 \times 10^{-27} \, 
{\rm g\, cm}^{-3}$, $v_{\rm gal}= 1100$ km s$^{-1}$, and assuming 
$\Sigma_{\rm gas}= \int_{-Z_0}^{+Z_0} \rho_{\rm ISM} dz \approx 1.2 \times 
10^{-3}\, {\rm g}\, {\rm cm}^{-2}$ (for $Z_0 = 100$ pc, and $ \rho_{\rm ISM}= 
2\times 10^{-24} \, {\rm g \, cm}^{-3}$), $v_{\rm rot} = 250$ km s$^{-1}$, 
we find 
$R_{\rm strip}= 6.2 \times 10^{21}$ cm. This means that all the gas 
down to a radius of 2 kpc will be stripped by the ram pressure, implying that 
the galaxy would lose more than 99\% of its gas, if $\Sigma_{\rm gas}$
were constant.  Even if we consider outer orbits, where $\rho_{\rm ICM}$ is
more than three times  lower, or partial
replenishment of the ISM from winds and supernovae (at a rate of $0.1 -
1 \, {\rm M_{\sun}/
yr}$) if the galaxy is actively forming stars, we still expect that gas
outside of $\sim$ 10 kpc will be stripped.

\noindent $\bullet$  As the gas streaming behind the bow shock is supersonic with
respect to the H{\sc i} gas, it should heat up the stripped gas to X-ray
temperatures as observed.  However, the gas in the tail will have a
tendency to be cooler than the ambient ICM if stripping is substantial
\citep[cf.][]{st99}, which is indeed seen in case of the X-ray
ridge in the {IC1262} group.

\noindent
$\bullet$  The estimated radiative cooling time of the X-ray ridge of
$\sim 10^9$ yr, and a relative velocity 
of  $v_{\rm gal}= 1100$ km s$^{-1}$ indicate that 
the X-ray bright tail might be observable over a 
distance of $\sim 1.1$ Mpc. This is comfortably larger than the length of 
the ridge, and therefore it is likely that the galaxy is for the first time 
plunging through the centre of the group.

In conclusion, there seems to be circumstantial evidence that the
conspicuous X-ray ridge could be the result of ram pressure stripping
of a member galaxy that is crossing the centre of the group 
on a nearly radial orbit, as expected from a late type galaxy \citep[][and
references therein]{BivianoKatgert}. 
However, we cannot easily explain
the full perturbation with just a trail of the orbital
motion of IC 1263, which will require a deeper understanding of the 
detailed dynamics
of the 
system.   

\section{Conclusions}

Sensitive Chandra and XMM-Newton observations of the IC1262 group have been
used to characterize the perturbation at the center of an otherwise extended,
regular emission from the hot gas in a seemingly relaxed, massive group of
galaxies. 
The sharp and narrow 
filamentary structure that runs in a NS direction, east of the central galaxy,
plus the loop closer to IC1262, 
are cooler than the surrounding medium, with steep drops
on either side, and discontinuities of $\sim 2\times$ in surface
brightness.  The total length of this apparently coherent structure
reaches up to a few hundred kpc. 
The nature of this complex feature is still elusive, possibly for lack of
supporting evidence at different wavelengths, that would highlight for example  the dust
content, ionized gas distribution  and detailed radio morphology of the system.  
We have considered possible interpretations, in light of similar
structures observed and studied in other groups or clusters of galaxies,
but all analogies are coupled with significant differences, mostly in
the size of the structure also relative to the sources of perturbation,
that make the understanding of this system far from exhaustive.
Its cooler temperature and the lack of evidence of a significant bulk
motion rule
out the possibility that this is a shock front.   We have found two
possible processes that could explain part of the observed properties
and might both be present:

\noindent $\bullet$ Ram pressure stripping of the
bright spiral IC1263, located $4'$ N of IC1262, is a promising process,
and would explain the length of the filament and its temperature.
With the current data it is hard to understand whether IC1263 is
bound to the group or is a newcomer.  In the former case, the galaxy
is most likely on a radial orbit, possibly
approaching the central dense core, where stripping is more efficient
\citep[see also][]{Pryor}, for the first time.  

\noindent $\bullet$   A radio
source is likely to significantly affect and modify the morphology and
characteristics of the hot ambient gas. Although the current data are of
limited quality, they indicate a low luminosity radio source, that cannot
compete in strength with the more spectacular examples in the literature.
Nonetheless, we cannot discard the possibility that the source was more
powerful in the past, and that the current perturbation is but a relic
of a since faded outburst.  A more thorough investigation of the radio
properties of this system, with deeper and higher resolution observations,
are needed to investigate this matter further.

\begin{acknowledgements} 
This research has made use of SAOImage DS9, developed by Smithsonian
Astrophysical Observatory  and  of the NASA/IPAC Extragalactic Database
(NED) which is operated by the Jet Propulsion Laboratory, California
Institute of Technology, under contract with the National Aeronautics
and Space Administration.  GT thanks
the Max-Planck-Institut f\"ur extraterrestrische Physik for the kind
hospitality and the stimulating environment during the paper preparation.
GT and AW acknowledge 
partial financial support from the Agenzia Spaziale Italiana
under contract ASI-INAF I/023/05/0.  \end{acknowledgements}

\bibliographystyle{}

\end{document}